\documentclass[12pt]{article}
\usepackage{amsmath}
\usepackage{graphicx,psfrag,epsf}
\usepackage{enumerate}
\usepackage{natbib}
\usepackage{url} 

\newcommand{\blind}{0}

\begin{document}

\def\spacingset#1{\renewcommand{\baselinestretch}%
{#1}\small\normalsize} \spacingset{1}

\def\hang{\hangindent\parindent}
\def\rf{\par\noindent\hang}

\newtheorem{theorem}{Theorem}
\newtheorem{lemma}{Lemma}
\newtheorem{proposition}{Proposition}

\newtheorem{definition}{Definition}

\newtheorem{example}{Example}
\newcommand{\bs}{\boldsymbol}

\newcommand{\bx}{x}
\newcommand{\bX}{x}
\newcommand{\bmu}{\boldsymbol{\mu}}

\newcommand{\SSB}{\text{\small SSB}}
\newcommand{\SSW}{\text{\small SSW}}
\newcommand{\SSG}{\text{\small SSG}}
\newcommand{\CP}{\text{\small CP}}
\newcommand{\CPK}{\text{\small CPK}}
\newcommand{\T}{\text{\small T}}


\if0\blind
{
  \title{\bf The impact of a Hausman pretest on the coverage probability and expected length of confidence intervals}
  \author{Paul Kabaila \thanks{
    \noindent  Corresponding author. Department of Mathematics and Statistics,
La Trobe University, Victoria 3086, Australia. Tel.: +61 3 9479 2594; fax +61 3 9479 2466.
{\sl E-mail address:} P.Kabaila@latrobe.edu.au.}\hspace{.2cm},
    Rheanna Mainzer
    and
     Davide Farchione \\
    Department of Mathematics and Statistics, La Trobe University}
  \maketitle
} \fi

\if1\blind
{
  \bigskip
  \bigskip
  \bigskip
  \begin{center}
    {\LARGE\bf Title}
\end{center}
  \medskip
} \fi

\bigskip
\begin{abstract}
In the analysis of clustered and longitudinal data, which includes a covariate that varies both between and within clusters (e.g. time-varying covariate in longitudinal data), a Hausman pretest is commonly used to decide whether subsequent inference is made using the linear random intercept model or the fixed effects model.
We assess the effect of this
pretest on the coverage probability and expected length of a confidence interval for the slope parameter.
Our results show that for the small levels of significance of the Hausman pretest commonly used in applications,
the minimum coverage probability of this confidence interval can
be far below nominal.  Furthermore, the expected length of this confidence interval is, on average, larger than the expected length of a confidence interval for the slope parameter based on the fixed effects model with the same minimum coverage.
\end{abstract}

\noindent%
\vfill

\newpage
\spacingset{1.45} 
\section{Introduction}


The linear random intercept model is commonly used in the analysis of clustered and longitudinal data. In clustered data the response variable is measured once on a unit where each unit is nested within a particular cluster of units. For example, analyzing the reading test score of children which are nested within classrooms (clusters). Longitudinal data, which can also be viewed as clustered data (see Rabe-Hesketh and Skrondal (2012, p. 227)), is where the response variable is measured at different time points for each subject and where the measurements across time are nested within each subject (cluster). For example, analyzing the weights of individuals over time where the weight measurements across time are nested within individuals (clusters). When including a covariate in the linear random intercept model that varies both between and within clusters (e.g. time-varying covariate in longitudinal data) a preliminary Hausman (1978) test is commonly used to test the assumption of no correlation between the random intercept and covariate. If the Hausman pretest rejects the null hypothesis of no correlation between the random intercept and covariate then the fixed effects model is chosen for subsequent inference, otherwise the linear random intercept model is chosen. This preliminary model selection procedure has been widely used in econometrics (see e.g. Wooldridge, 2002 and Baltagi, 2005) and has also been adopted in other areas such as medical statistics, see e.g. Gardiner et al. (2009) and Mann et al. (2004). The Hausman pretest has also been implemented in popular statistical computer programmes including SAS, Stata, eViews and R, see Ajmani (2009, Chapter 7.5.3), Rabe-Hesketh and Skrondal (2012, Chapter 3.7.6), Griffiths et al. (2012, Chapter 10.4) and Croissant and Millo (2008), respectively.



The two-stage procedure widely used in the analysis of clustered and longitudinal data is as follows.
In the first stage, the Hausman pretest is used to decide
whether subsequent inference is made using the random intercept model or the fixed effects model (see e.g. Ebbes et al., 2004 and Jackowicz et al., 2013). The second stage is that the inference
of interest is carried out assuming that the model chosen in the first stage had been given to us \textit{a priori}, as the true model.
Guggenberger (2010) considers this two-stage procedure when the inference of interest is a hypothesis test about the slope
parameter.
He provides both a local asymptotic analysis of the size of this test and a finite sample analysis (via simulations) of the probability of Type I error.

We consider the case
that the inference of interest is a confidence interval for the slope parameter.
Kabaila et al., 2015 state 3 new finite sample results (Theorems \ref{thm: CovProb_DependOnPsi}, \ref{thm: eveness} and \ref{known coverage} of the present paper)
on the effect of a Hausman pretest on the coverage probability of the
confidence interval for this parameter. They also provide outline proofs of these results and a brief initial analysis of the
coverage properties of this confidence interval. These coverage properties are estimated by simulation.
In the present paper, we describe how the third of these results can be used
to substantially improve the efficiency of these simulations through the use of variance reduction by control variates.

We compare the expected length of the confidence interval resulting from the two-stage procedure with the expected length of the confidence interval based on the fixed effects model, where the latter interval is adjusted to have the same minimum coverage as the former interval.
The quantity used for this comparison is the scaled expected length, defined as the expected length of the interval resulting from the two-stage procedure divided by the expected length of the adjusted interval from the fixed effects model.
The scaled expected length provides an insight into how useful the Hausman pretest is in this context.
In the present paper, we provide 4 new finite sample theorems (Theorems \ref{thm: cov_standard_int}, \ref{thm: SEL}, \ref{thm: SEL_DependOnPsi} and \ref{thm: eveness_SEL})
on the scaled expected length.
The scaled expected length is estimated by simulation. We describe how
to substantially improve the efficiency of these simulations through the use of variance reduction by control variates.

The results that we present make it easy to assess, for a wide variety of circumstances, the effect of the Hausman pretest on the coverage properties and scaled expected length of the confidence interval for the slope parameter.
Our results show that when the usual small nominal level of significance for the Hausman pretest is used, the two-stage procedure can result in a confidence interval with minimum coverage probability far below nominal. However, if the nominal level of significance is increased to 50\% then the minimum coverage probability is much closer to the nominal coverage
(see Figures 1, 2 and 3).
We also show that, in terms of expected length, the confidence interval resulting from the two-stage procedure is consistently outperformed by the confidence interval based on the fixed effects model, regardless of the nominal level of significance chosen for the Hausman pretest (see Table 1).
The results presented in this paper were computed using programs written in the R programming language, which will be made available in a convenient R package.

In Section 2, we consider the practical situation that the
random error and random effect variances
are estimated from
the data. We consider three estimators of these variances: the usual unbiased estimators, the maximum likelihood estimators of Hsiao (1986) and the estimators of Wooldridge (2002).
The coverage probability and the scaled expected length
of the confidence interval resulting from the two-stage procedure are determined by 4 known quantities and 5 unknown
parameters. The known quantities are the number of individuals, the number of time points (for the longitudinal data case), the nominal significance level of the Hausman pretest and the nominal coverage probability of this confidence interval. The unknown parameters are the random error variance, the random effect variance, the variance of the covariate (or time-varying covariate in the case of longitudinal data), a scalar parameter that determines the correlation matrix of the covariates and a non-exogeneity
parameter.

If, for given values of the 4 known quantities, we wish to assess the dependence of the coverage probability
and the scaled expected length
of the confidence interval resulting from the two-stage procedure on the 5 unknown parameters then we might consider, say, five values for each of these unknown parameters, leading to 3125 parameter combinations. Apart from the daunting task of summarizing so many results,
 it is possible that one might miss important values of the unknown parameters, such as values for which the
coverage probability is particularly low or the scaled expected length is particularly large.

Theorems \ref{thm: CovProb_DependOnPsi} and \ref{thm: SEL_DependOnPsi} state that, apart from the known quantities, the coverage probability and scaled expected length
are actually determined by only 3 unknown
parameters, including the non-exogeneity parameter. If we compute the minimum coverage probability with respect to the non-exogeneity parameter then
we have only 2 unknown parameters and our assessment of the coverage properties of the confidence interval resulting from the two-stage procedure is
greatly simplified. Theorems \ref{thm: eveness} and \ref{thm: eveness_SEL} state that the coverage probability and scaled expected length are even functions of the non-exogeneity parameter, so that
computation time is halved.  We also propose a scaling of the non-exogeneity parameter that takes account of the sample size. In effect, this scaling reduces the number of known quantities that determine this coverage probability
 from 4 to 3.

In Section 3, we consider the coverage probability of the confidence interval
resulting from the two-stage procedure when the random error and random effect variances are assumed to be known.
Theorem \ref{known coverage} shows that the conditional coverage probability of the confidence interval resulting from the two-stage procedure
 can be found exactly by the evaluation of
the bivariate normal cumulative distribution function. This theorem is important because it is used to reduce the variance of the simulation based estimators of the coverage probability and scaled expected length
of the confidence interval resulting from the two-stage procedure (when random error and random effect variances are estimated). As we show in Section 4 this variance reduction is achieved by using control variates.

\section{
The model and the practical two-stage procedure (random error and random effect variances are \newline estimated)}

We focus on the case of longitudinal data for which $i$ denotes the individual $(i = 1, \dots, N)$ and $t$ denotes the time period $(t = 1, \dots, T)$.  By interpreting $i$ as the cluster index and $t$ as the unit of analysis, our results also apply to the analysis of clustered data.
Part of the following description of the model and two-stage procedure is taken from Kabaila et al., 2015.
Let $y_{it}$ and $x_{it}$ denote the response variable and the time-varying covariate, respectively, for the $i$'th individual 
at time $t$
. Suppose that
\begin{equation}
\label{model}
y_{it} = a + \beta x_{it} + \mu_i + \varepsilon_{it},
\end{equation}
where the $\varepsilon_{it}$'s and the $(\mu_i, x_{i1}, \dots, x_{iT})$'s are independent, the $\varepsilon_{it}$'s are i.i.d. $N(0, \sigma_{\varepsilon}^2)$ and the $\mu_i$'s are i.i.d. $N(0, \sigma_{\mu}^2)$. We call $\beta$ the slope parameter, $\sigma_{\varepsilon}^2$ the error variance
and $\sigma_{\mu}^2$ the random effect variance. The $\varepsilon_{it}$'s and the
$\mu_i$'s are unobserved.
Suppose that the parameter of interest is the slope parameter $\beta$ and that the inference of interest is a confidence interval for $\beta$.

Also suppose that the $(\mu_i, x_{i1}, \dots, x_{iT})$'s are
i.i.d. multivariate normally distributed with zero mean and covariance matrix
\begin{equation}
\label{CovMatrixMuiXit}
\begin{bmatrix}
\sigma^2_{\mu} & \widetilde{\tau}\sigma_{\mu}\sigma_x e^{\prime} \\
\widetilde{\tau}\sigma_{\mu} \sigma_x e & \sigma^2_x G
\end{bmatrix},
\end{equation}
where $e$ is a $T$ vector of 1's, $G$  is a $T \times T$ matrix with $1$'s on the diagonal and $\rho$ on the off-diagonal (compound symmetry), and $\tilde{\tau}$ is a parameter that measures the dependence between $\mu_{i}$ and $(x_{i1}, \dots, x_{iT})$.
We define the ``non-exogeneity parameter'' $\tau = \widetilde{\tau} \big(T/(1+(T-1)\rho) \big)^{1/2}$ and note that
it is a correlation, so $\tau \in (-1, 1)$.
If $\tau = 0$ then the
$x_{it}$'s are exogenous variables.
Let $x = (x_{11}, \dots, x_{1T}, x_{21}, \dots, x_{2T}, \dots, x_{N1}, \dots, x_{NT})$.
In simulations, we generate $x$ as follows. Suppose that
$u_{11}, \dots, u_{1T}, \dots, u_{N1}, \dots, u_{NT}, v_1, \dots, v_N$ are i.i.d. $N(0, 1)$.
Also suppose that these random variables and the $\varepsilon_{it}$'s are independent.
Let $x_{it} / \sigma_x = (1 - \rho)^{1/2} \, u_{it} + \rho^{1/2} \, v_i$ for $i=1,\dots,N$
and $t=1,\dots,T$. 
The distribution of $\mu_i$ conditional on $(x_{i1}, \dots, x_{iT})$, found from \eqref{CovMatrixMuiXit},
is used to generate $(\mu_1, \dots, \mu_N)$.
Let
$u = (u_{11}, \dots, u_{1T}, \dots, u_{N1}, \dots, u_{NT}, v_1, \dots, v_N)$.

Assume, for the moment, that $\sigma_{\varepsilon}$ and $\sigma_{\mu}$ are known.
 When $\tau = 0$, a confidence interval
for $\beta$ may be found as follows. Let
$\psi = \sigma_{\mu}  / \sigma_{\varepsilon}$.
Condition on $x$ and use the GLS estimator
$\widehat{\beta}(\psi)$ of $\beta$. Let $z_c = \Phi^{-1}(c)$, where $\Phi$ denotes the $N(0,1)$ cdf.
Define the following confidence interval for $\beta$
\begin{equation*}
I(\psi) = \left[\widehat{\beta}(\psi) - z_{1-\alpha/2} \big(\text{Var}_0(\widehat{\beta}(\psi) \, | \, x)\big)^{1/2}, \, \widehat{\beta}(\psi)
+ z_{1-\alpha/2} \big(\text{Var}_0(\widehat{\beta}(\psi) \, | \, x)\big)^{1/2} \right],
\end{equation*}
where $\text{Var}_0(\widehat{\beta}(\psi) \, | \, x)$ denotes the variance of $\widehat{\beta}(\psi)$, conditional on $x$ when $\tau = 0$.
The confidence interval $I(\psi)$ has coverage probability $1-\alpha$ when $\tau = 0$.
Averaging \eqref{model} over $t=1, \dots, T$ for each $i=1, \dots, N$ we obtain
\begin{equation}
\label{average_over_t_model}
\overline{y}_i = a + \beta \overline{x}_i + \mu_i + \overline{\varepsilon}_i,
\end{equation}
where
\begin{equation*}
\overline{y}_i = \frac{1}{T} \sum_{t=1}^T y_{it} \ , \ \ \ \overline{x}_i = \frac{1}{T} \sum_{t=1}^T x_{it}
\ \ \ \text{and} \ \ \ \ \overline{\varepsilon}_i = \frac{1}{T} \sum_{t=1}^T \varepsilon_{it}.
\end{equation*}
This model is called the {\sl between effects model}.
When $\tau = 0$, an alternative estimator of
$\beta$ is $\widetilde{\beta}_B$, the OLS estimator based on the model \eqref{average_over_t_model}, when we condition on $x$. This estimator does not require a knowledge of $\psi$.
Subtracting
\eqref{average_over_t_model} from \eqref{model}, we obtain
\begin{equation}
\label{consistent_est_beta_model}
y_{it} - \overline{y}_i = \beta (x_{it} - \overline{x}_i) + (\varepsilon_{it} - \overline{\varepsilon}_i).
\end{equation}
This model is called the {\sl fixed effects model}. We estimate $\beta$
by $\widetilde{\beta}_W$, the OLS estimator based on this model.
Define the following confidence interval for $\beta$
\begin{equation*}
J(\sigma_{\varepsilon}) = \left [\widetilde{\beta}_W - z_{1-\alpha/2} \big(\text{Var}(\widetilde{\beta}_W \, | \, x)\big)^{1/2},
\widetilde{\beta}_W +
z_{1-\alpha/2} \big(\text{Var}(\widetilde{\beta}_W \, | \, x)\big)^{1/2} \right ],
\end{equation*}
where $\text{Var}(\widetilde{\beta}_W \, | \, x)$ denotes
the variance of $\widetilde{\beta}_W$, conditional on $x$.
Irrespective of the value of $\tau$,
the confidence interval $J(\sigma_{\varepsilon})$
has coverage probability $1-\alpha$.
In practice, we do not know whether or not $\tau = 0$. The usual procedure is to use a Hausman pretest to test $H_0: \tau = 0$ against $H_a: \tau \not= 0$. We consider this pretest, based on the test statistic
\begin{equation}
\label{Def_H}
H(\sigma_{\varepsilon},\sigma_{\mu}) =  \frac{(\widetilde{\beta}_W-\widetilde{\beta}_B)^2}{\text{Var}(\widetilde{\beta}_W \, | \, x) + \text{Var}_0(\widetilde{\beta}_B \, | \, x)},
\end{equation}
where $\text{Var}_0(\widetilde{\beta}_B \, | \, x)$ denotes the variance of
$\widetilde{\beta}_B$ conditional on $x$ and assuming that $\tau = 0$.
This test statistic has a $\chi^2_1$ distribution under $H_0$.
Suppose that we accept $H_0$ if
$H(\sigma_{\varepsilon},\sigma_{\mu})
\le z^2_{1 - \widetilde{\alpha}/2}$; otherwise we reject $H_0$.
Note that $\widetilde{\alpha}$ is the level of significance of this test.
We now describe the two-stage procedure.
If $H_0$ is accepted then we use the confidence interval $I(\psi)$; otherwise we use the confidence interval $J(\sigma_{\varepsilon})$. Let $K(\sigma_{\varepsilon},\sigma_{\mu})$ denote the confidence interval, with nominal coverage
$1-\alpha$, that results from this two-stage procedure.

Of course, in practice, $\sigma_{\varepsilon}$ and $\sigma_{\mu}$ are not known and need to be estimated. So, in practice, the Hausman pretest is based on the test statistic $H(\widehat{\sigma}_{\varepsilon}, \widehat{\sigma}_{\mu})$ and the two-stage procedure results in the confidence interval $K(\widehat{\sigma}_{\varepsilon},\widehat{\sigma}_{\mu})$ where $\widehat{\sigma}_{\varepsilon}$ and $\widehat{\sigma}_{\mu}$ denote estimators of $\sigma_{\varepsilon}$ and $\sigma_{\mu}$, respectively.

\subsection{The coverage probability of the confidence interval resulting from the two-stage procedure}

The coverage probability of the confidence interval constructed from this two-stage procedure is $P(\beta \in K(\widehat{\sigma}_{\varepsilon}, \widehat{\sigma}_{\mu} ))$. The following two theorems (stated by Kabaila et al. 2015) give properties of
this coverage probability.

\begin{theorem}
\label{thm: CovProb_DependOnPsi}


\smallskip

\noindent For $(\widehat{\sigma}_{\varepsilon},\widehat{\sigma}_{\mu})$ any of the
pairs of estimators listed in Appendix B, the unconditional coverage probability $P(\beta \in K(\widehat{\sigma}_{\varepsilon}, \widehat{\sigma}_{\mu} ))$ is determined by $N$ (the number of individuals), $T$ (the number of time points), $\widetilde{\alpha}$ (the nominal significance level of the Hausman pretest), $1 - \alpha$ (the nominal coverage probability), $\psi$ (the ratio $\sigma_{\mu}/\sigma_{\varepsilon}$), $\rho$ (the parameter that determines $G$) and $\tau$ (the non-exogeneity parameter).  Given these quantities, the coverage probability does not depend on either $\sigma^2_{\varepsilon}$ (the variance of the random error) or $\sigma^2_{\mu}$ (the variance of the random effect) or $\sigma^2_x$ (the variance of the time-varying covariate $x_{it}$).

\end{theorem}


\begin{theorem}
\label{thm: eveness}
Suppose that $N$, $T$, $\widetilde{\alpha}$, $1-\alpha$, $\psi$ and $\rho$ are fixed.
When $\sigma_{\varepsilon}$ and  $\sigma_{\mu}$ are replaced by any of the pairs of estimators listed in Appendix B,
the unconditional coverage probability $P(\beta \in K(\widehat{\sigma}_{\varepsilon}, \widehat{\sigma}_{\mu} ))$
is an even function of $\tau \in (-1,1)$.
\end{theorem}

Outline proofs of these results are provided by Kabaila et al., 2015.
Detailed proofs are provided in the supplementary material for the present paper.
A remarkable feature of the proofs of these theorems is that they are carried out without relying on a simple expression for the
coverage probability $P(\beta \in K(\widehat{\sigma}_{\varepsilon}, \widehat{\sigma}_{\mu} ))$.
We use simulations to compute $P(\beta \in K(\widehat{\sigma}_{\varepsilon}, \widehat{\sigma}_{\mu} ))$, employing
variance reduction by control variates, as described in Section 4.
Using Theorem \ref{thm: eveness}, we only need to consider $\tau$ in the interval $[0, 1)$, which means that we have reduced the number of simulations needed to estimate the coverage probability function (or its minimum) by half.

We now examine the influence that the nominal level of significance $\widetilde{\alpha}$ of the Hausman pretest has on the coverage probability function $P(\beta \in K(\widehat{\sigma}_{\varepsilon}, \widehat{\sigma}_{\mu} ))$.
Suppose that
$\widehat{\sigma}_{\varepsilon}$ and $\widehat{\sigma}_{\mu}$
are the usual unbiased estimators of $\sigma_{\varepsilon}$ and $\sigma_{\mu}$, respectively (described in Appendix B).
 Consider $\rho = 0.3$, $N = 100$, $T = 3$, $\psi =  \sigma_{\mu} / \sigma_{\varepsilon} = 1/3$ and the
nominal coverage probability $1-\alpha = 0.95$.
In practice, it is common to use a small value of $\widetilde{\alpha}$, such as 0.05 or 0.01.
As noted by Guggenberger (2010), examples of practical applications that have used a small $\widetilde{\alpha}$ for the Hausman pretest are provided by Gaynor et al. (2005, p.245) and Bedard and Deschenes (2006, p.189).
Figure 1 presents graphs of the coverage probability
$P(\beta \in K(\widehat{\sigma}_{\varepsilon}, \widehat{\sigma}_{\mu} ))$,
considered as a function of $\lambda = N^{1/2} \tau$.
Each graph is computed using the variance reduction method and the common random numbers (to produce smoother graphs)
described in Section 4.
The number of simulation runs used to compute each graph is $M = 20000$.
The bottom (with circle points) graph is for nominal significance level $\widetilde{\alpha} = 0.05$ of the Hausman pretest.
This graph falls well below the nominal
coverage for a wide interval of values of $\lambda$, with the minimum of the coverage probability approximately equal to 0.75.
Suppose that we choose the significance level of the Hausman pretest to be quite large, say $\widetilde{\alpha}=0.50$.  Now the Hausman pretest is  more likely to reject the null hypothesis that $\tau =0$ and therefore more likely to choose the fixed effects model for the construction of the
confidence interval.
The top (with triangle points) graph is for nominal significance level $\widetilde{\alpha} = 0.50$ of the Hausman pretest.
Although this graph is still below the nominal coverage, there has been a large improvement.
%

\begin{figure}[h!]
 \centering

  \includegraphics[scale=0.7]{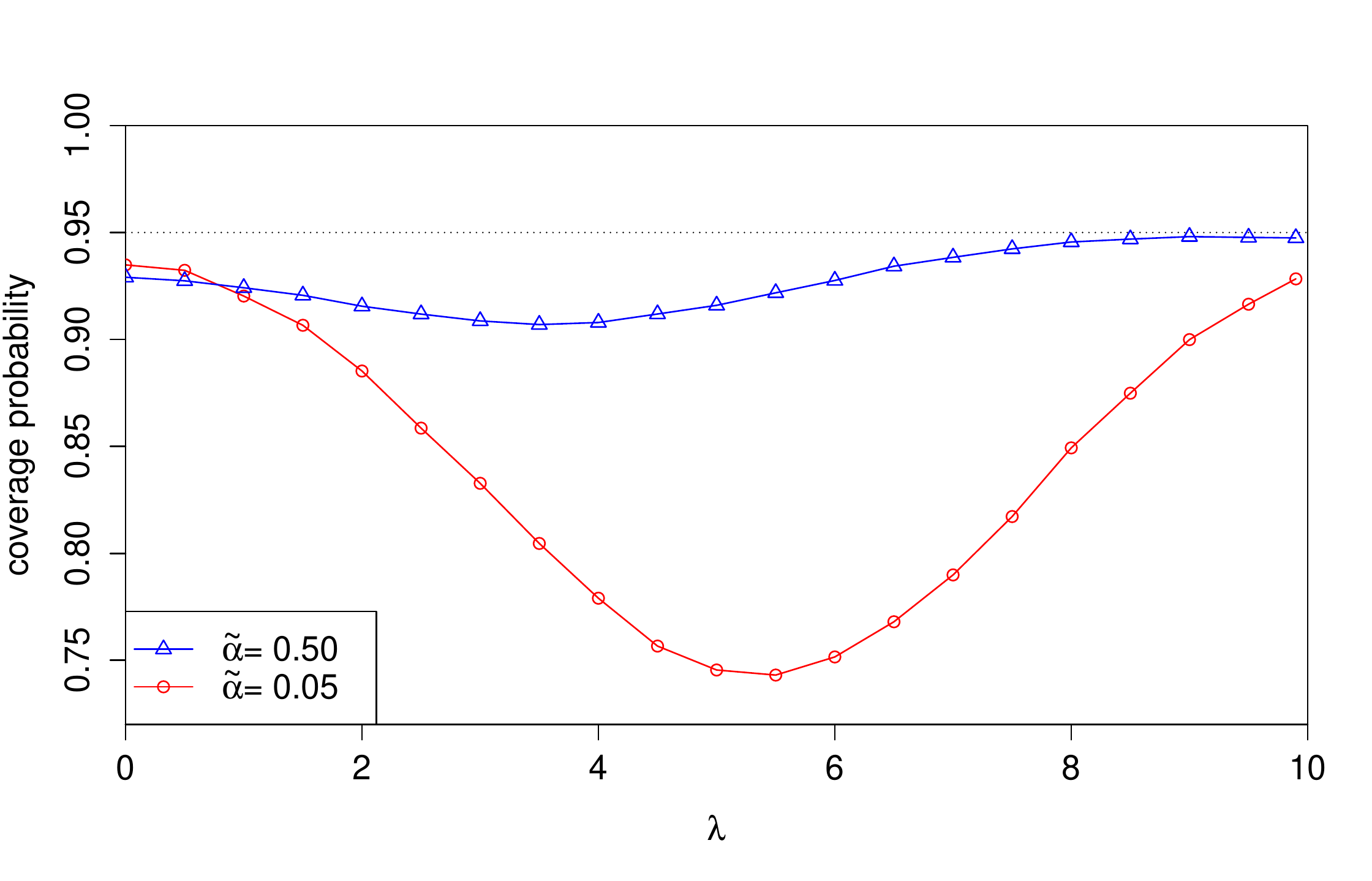}
 \caption{Graphs of the coverage probability functions of the confidence interval resulting from the two-stage procedure, for nominal coverage probability 0.95.
 Here $\lambda = N^{1/2}\tau$, where $\tau$ is the non-exogeneity parameter, $\rho= 0.3$, $N = 100$, $T = 3$ and $\psi = 1/3$.
Two nominal levels of significance, $\widetilde{\alpha}=0.05$ and $\widetilde{\alpha}=0.5$, of the Hausman pretest are considered.}
 \label{fig:Fig1}
\end{figure}

%
%

As noted in the Introduction and in Section 2, if we compute the minimum over $\tau$ of the coverage probability
$P(\beta \in K(\widehat{\sigma}_{\varepsilon}, \widehat{\sigma}_{\mu}))$ then we are left with only two unknown parameters, $\psi$ and $\rho$.
If we fix $\psi$ then the minimum coverage depends only on $\rho$, where $\rho \in (-1, 1)$, as it is a correlation.
Suppose that
$\widehat{\sigma}_{\varepsilon}$ and $\widehat{\sigma}_{\mu}$
are the usual unbiased estimators of $\sigma_{\varepsilon}$ and $\sigma_{\mu}$, respectively.
Suppose that $N = 100$, $T = 3$, $\psi =  \sigma_{\mu} / \sigma_{\varepsilon} = 1/3$ and the
nominal coverage probability $1-\alpha = 0.95$.
Figure 2 presents graphs of the coverage probability
$P(\beta \in K(\widehat{\sigma}_{\varepsilon}, \widehat{\sigma}_{\mu} ))$, minimized over $\tau$,
considered as a function of $\rho$ ($\rho \ge 0$).
It can be seen that the coverage probability, minimized over $\tau$, is a decreasing function of $\rho$.  This is because as $\rho$ increases, $x_{it} - \overline{x}_i$ becomes closer to 0 for $t = 1, \dots, T$ ($i=1, \dots, N$), causing $\widetilde{\beta}_W$ to become a very inaccurate estimator of $\beta$.
This then causes the test statistic $H(\widehat{\sigma}_{\varepsilon}, \widehat{\sigma}_{\mu})$ to have very little power to detect non-zero values of $\tau$.
Each estimate of the minimum coverage is found using the common random numbers and
 the variance reduction method described in Section 4.
Similarly to Figure 1, we see a vast improvement in the minimum coverage by letting $\widetilde{\alpha} = 0.50$ rather than choosing $\widetilde{\alpha}$ to be the commonly used, smaller value 0.05.
%

\begin{figure}[h!]
 \centering
 \includegraphics[scale=0.7]{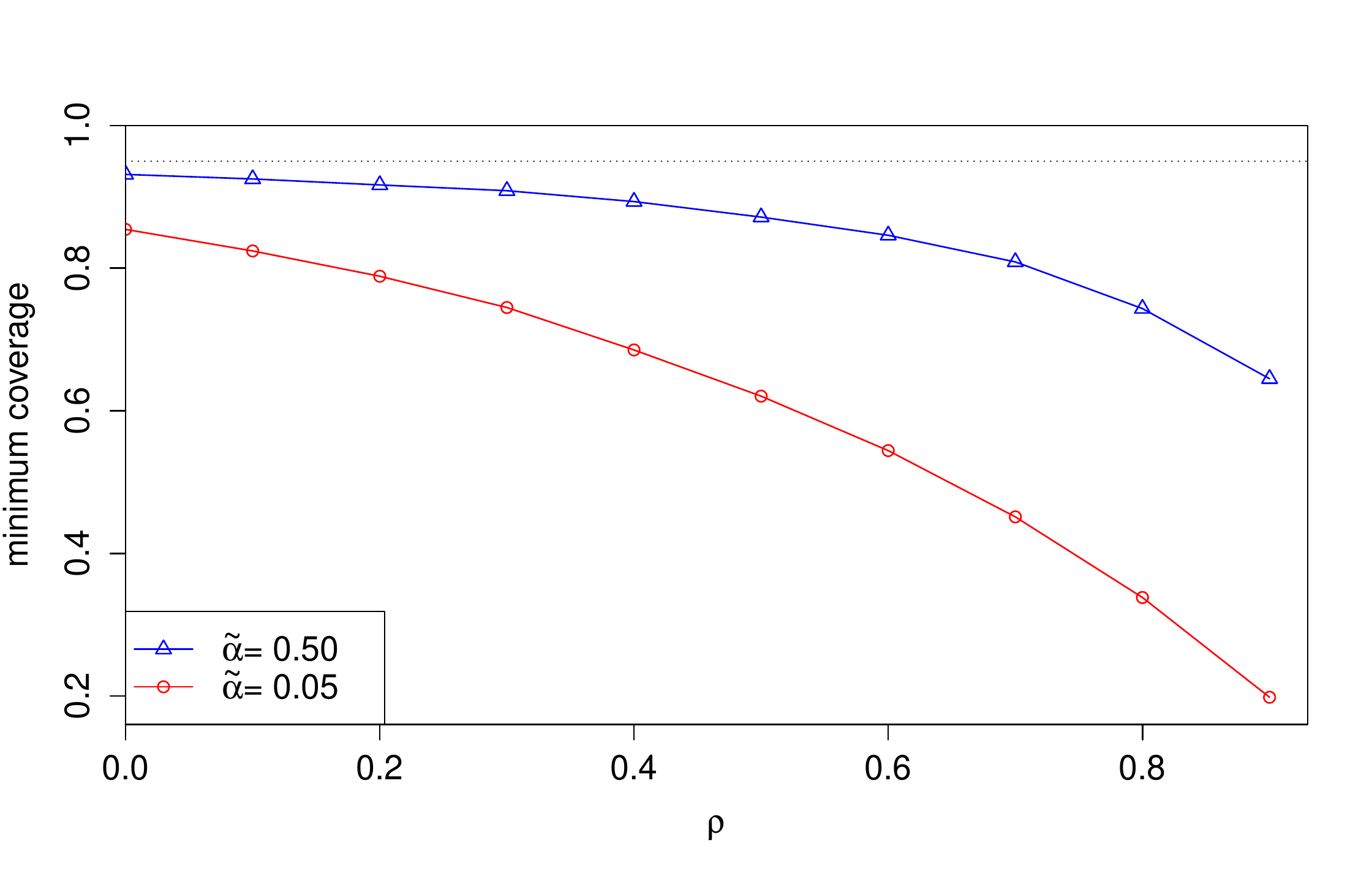}
 \caption{Graphs of the coverage probability functions, minimized over the non-exogeneity parameter $\tau$, of the confidence interval resulting from the two-stage procedure for nominal coverage probability 0.95.
 This minimum coverage is considered as a function of $\rho$, the parameter that determines $G$.
 Two nominal levels of significance, $\widetilde{\alpha}=0.05$ and $\widetilde{\alpha}=0.5$, of the Hausman pretest are considered.
Here $N = 100$, $T = 3$ and $\psi = 1/3$.}
 \label{fig:Fig3}
\end{figure}

In practice, $\psi$ is not known and must be estimated from the data.  However, it is commonly observed in practice that $\rho$ takes small to moderate values, and not values close to 1.  This suggests that we fix $\rho$ and plot the graph of the coverage probability
$P(\beta \in K(\widehat{\sigma}_{\varepsilon}, \widehat{\sigma}_{\mu} ))$, minimized over $\tau$, as a function of $\psi$.
Suppose that
$\widehat{\sigma}_{\varepsilon}$ and $\widehat{\sigma}_{\mu}$
are the usual unbiased estimators of $\sigma_{\varepsilon}$ and $\sigma_{\mu}$, respectively.
Consider $N = 100$, $T = 3$, the nominal
 coverage probability $1-\alpha = 0.95$ and $\rho = 0.4$.
Figure 3 presents graphs of the coverage probability
$P(\beta \in K(\widehat{\sigma}_{\varepsilon}, \widehat{\sigma}_{\mu} ))$, minimized over $\tau$,
as a function of $\psi$.
For nominal significance level $\widetilde{\alpha} = 0.05$ of the Hausman pretest, this minimized coverage probability
is far below the nominal coverage
for $\psi$ approximately equal to 0.2. However, for nominal significance level $\widetilde{\alpha} = 0.5$ of the Hausman pretest, we see
(once more) a dramatic
improvement in the minimum coverage probability.
%

\begin{figure}[h]
 \centering
 \includegraphics[scale=0.7]{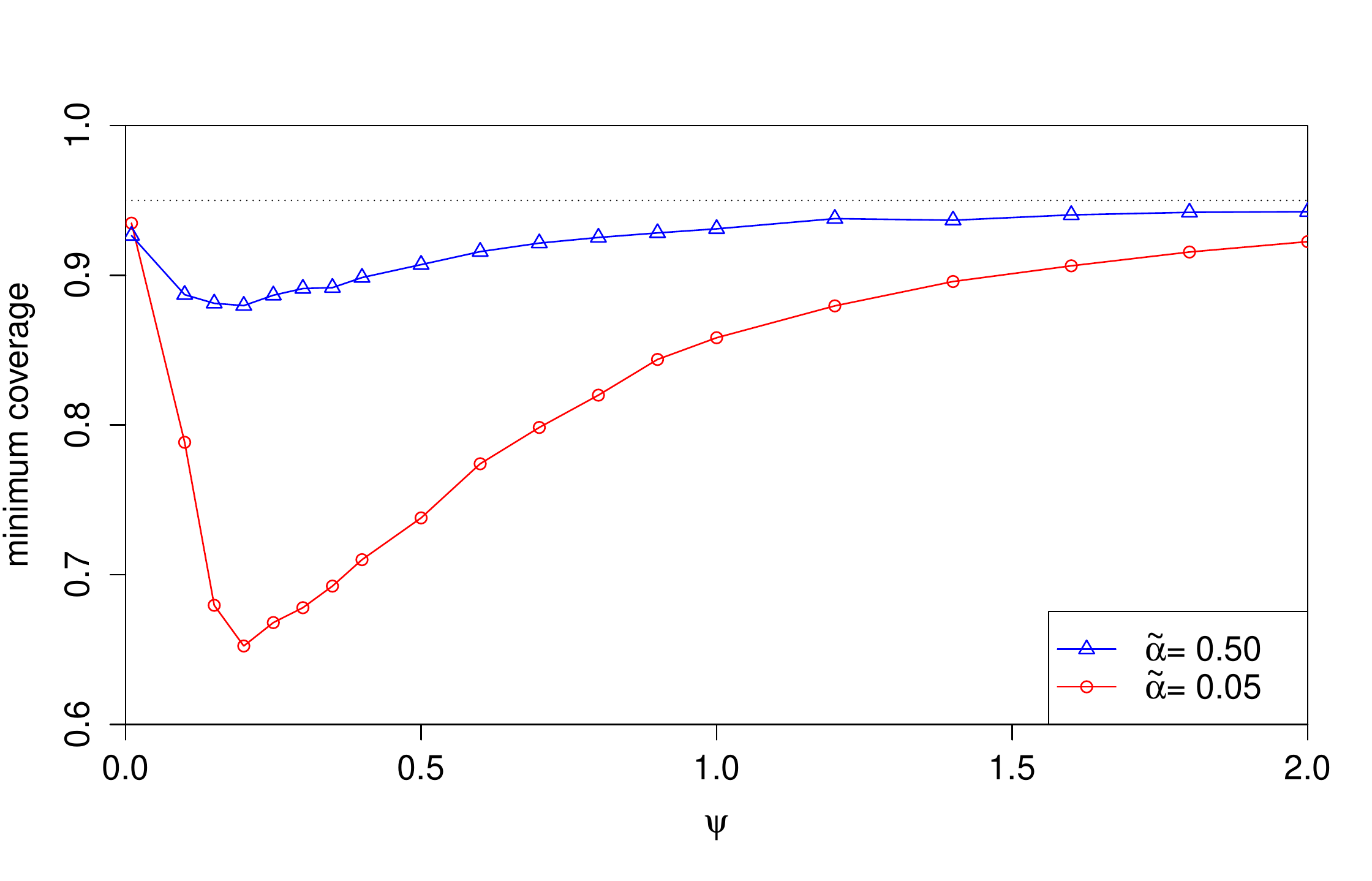}
 \caption{
Graphs of the coverage probability functions, minimized over the non-exogeneity parameter $\tau$, of the confidence interval resulting from the two-stage procedure for nominal coverage probability 0.95.
This minimum coverage is considered as a function of $\psi = (\text{random intercept standard deviation})/(\text{random error standard deviation})$,
where $N = 100$, $T = 3$ and $\rho = 0.4$.
Two nominal levels of significance, $\widetilde{\alpha}=0.05$ and $\widetilde{\alpha}=0.5$, of the Hausman pretest are considered.}
 \label{fig:Fig4}
\end{figure}

\subsection{Comparison of the two-stage interval with the adjusted \newline interval based on the fixed effects model}

The following notation is introduced to describe the expected length of the two-stage confidence interval.
Let $\overline{x} = (NT)^{-1}\sum_{i=1}^N\sum_{t=1}^Tx_{it}$, $\SSB = \sum_{i=1}^N(\overline{x}_i - \overline{x})^2$ (``sum of squares between") and $\SSW = \sum_{i=1}^N \sum_{t=1}^T (x_{it} - \overline{x}_i)^2$ (``sum of squares within").
Let $r(x) = \SSB/\SSW$ and $q(\psi, T) = \psi^2 + (1/T)$.  Now define $w = q(\psi, T) / (q(\psi, T) + r(x))$ and $\widehat{w} = q(\widehat{\psi}, T) / (q(\widehat{\psi}, T) + r(x))$, where $\widehat{\psi}=\widehat{\sigma}_{\mu} / \widehat{\sigma}_{\varepsilon}$.

We use the notation
\begin{equation*}
{\cal I}({\cal A}) =
\begin{cases}
1 &\text{if } {\cal A} \ \ \text{is true} \\
0 &\text{if } {\cal A} \ \ \text{is false}
\end{cases}
\end{equation*}
where ${\cal A}$ is an arbitrary statement.
Let $\mathcal{B}$ be the statement $ H(\widehat{\sigma}_{\varepsilon},\widehat{\sigma}_{\mu} ) \le z^2_{1-\widetilde{\alpha}/2} $.
The expected length of $K(\widehat{\sigma}_{\varepsilon}, \widehat{\sigma}_{\mu})$ is
\begin{equation*}
\label{expected_lengh_K}
2 z_{1-\alpha/2} \, E \left(\widehat{\sigma}_{\varepsilon} \, \left( \SSW \right)^{-1/2} \left( \widehat{w}^{1/2} {\cal I} \left( B \right) + {\cal I}\left( B^c \right) \right) \right).
\end{equation*}
Values of $\rho$ close to 1 are unlikely to be encountered in practice.  We therefore restrict attention to $\rho \in [0, \widetilde{\rho}]$ for some given $\widetilde{\rho} \in (0, 1)$.
Let $c_{\text{min}}$ denote the coverage probability of $K(\widehat{\sigma}_{\varepsilon}, \widehat{\sigma}_{\mu})$, minimized over $\tau \in (-1, 1)$, $\psi \in (0, \infty)$ and $\rho \in [0, \widetilde{\rho}]$.

Let
\begin{align}
\label{standard_int_known_sigma_varepsilon}
\begin{split}
J_c(\sigma_{\varepsilon})
&= \left [\widetilde{\beta}_W - \Phi^{-1}((c+1)/2) \big(\text{Var}(\widetilde{\beta}_W \, | \, x)\big)^{1/2},
\widetilde{\beta}_W +
\Phi^{-1}((c+1)/2) \big(\text{Var}(\widetilde{\beta}_W \, | \, x)\big)^{1/2} \right ]
\\
&= \left [\widetilde{\beta}_W - \Phi^{-1}((c+1)/2) \, \sigma_{\varepsilon}/(\SSW)^{1/2},
\widetilde{\beta}_W +
\Phi^{-1}((c+1)/2) \, \sigma_{\varepsilon}/(\SSW)^{1/2} \right ].
\end{split}
\end{align}
The following new theorem allows us to easily compute $P(\beta \in J_c(\widehat{\sigma}_{\varepsilon}))$
for any given $c \in (0,1)$.

\begin{theorem}
\label{thm: cov_standard_int}
$P(\beta \in J_c(\widehat{\sigma}_{\varepsilon}))$ does not depend on any unknown parameters.
\end{theorem}
The proof of Theorem \ref{thm: cov_standard_int} is provided in the supplementary material.
\noindent Define $c^*$ to be the value of $c$ such that
$P(\beta \in J_c(\widehat{\sigma}_{\varepsilon})) = c_{\text{min}}$. Let
 $J_{\text{adj}}(\widehat{\sigma}_{\varepsilon})$ be $J_c(\sigma_{\varepsilon})$ for $c = c^*$.
In other words, $J_{\text{adj}}(\widehat{\sigma}_{\varepsilon})$
is the confidence interval based on the fixed effects model, when $\sigma_{\varepsilon}$ is estimated from the data, that is adjusted to have the same minimum coverage as $K(\widehat{\sigma}_{\varepsilon}, \widehat{\sigma}_{\mu})$.
The expected length of $J_{\text{adj}}(\widehat{\sigma}_{\varepsilon})$ is
\begin{equation*}
2 \Phi^{-1} \left( \left( c^* + 1\right) / 2\right) \, E \left( \widehat{\sigma}_{\varepsilon} \, \left( \SSW \right)^{-1/2}  \right).
\end{equation*}
%

The scaled expected length of $K(\widehat{\sigma}_{\varepsilon}, \widehat{\sigma}_{\mu})$ is defined to be the expected length of $K(\widehat{\sigma}_{\varepsilon}, \widehat{\sigma}_{\mu})$ divided by the expected length of $J_{\text{adj}}(\widehat{\sigma}_{\varepsilon})$.  Let $\overline{u}_i = T^{-1} \sum_{t=1}^T u_{it}$. A useful expression for this scaled expected length is given by the following result.

\begin{theorem}
\label{thm: SEL}
The scaled expected length of $K(\widehat{\sigma}_{\varepsilon}, \widehat{\sigma}_{\mu})$ is equal to
\begin{equation}
\label{SEL}
\frac{ z_{1-\alpha/2}}{\Phi^{-1}((c^*+1)/2)} \frac{ E \left( (\widehat{\sigma}_{\varepsilon}/\sigma_{\varepsilon}) \left( \sum_{i=1}^N \sum_{t=1}^T (u_{it} - \overline{u}_i)^2 \right)^{-1/2} \left( \widehat{w}^{1/2} {\cal I}(\mathcal{B}) + {\cal I}(\mathcal{B}^c) \right) \right)}{E \left( (\widehat{\sigma}_{\varepsilon}/\sigma_{\varepsilon}) \left( \sum_{i=1}^N \sum_{t=1}^T (u_{it} - \overline{u}_i)^2 \right)^{-1/2} \right)}.
\end{equation}
\end{theorem}

A proof of this result is provided in the supplementary material.
If \eqref{SEL} is less than 1 then $K(\widehat{\sigma}_{\varepsilon}, \widehat{\sigma}_{\mu})$ is a shorter interval on average than $J_{\text{adj}}(\widehat{\sigma}_{\varepsilon})$.
The following two new theorems describe important properties of the scaled expected length.  Outline proofs of these theorems are given in Appendix C.
Detailed proofs are provided in the supplementary material.

\begin{theorem}
\label{thm: SEL_DependOnPsi}


\noindent For $(\widehat{\sigma}_{\varepsilon},\widehat{\sigma}_{\mu})$ any of the
pairs of estimators listed in Appendix B, the unconditional scaled expected length given in \eqref{SEL} is determined by $N$ (the number of individuals), $T$ (the number of time points), $\widetilde{\alpha}$ (the nominal significance level of the Hausman pretest), $1 - \alpha$ (the nominal coverage probability), $\psi$ (the ratio $\sigma_{\mu}/\sigma_{\varepsilon}$), $\rho$ (the parameter that determines $G$) and $\tau$ (the non-exogeneity parameter).  Given these quantities, the scaled expected length does not depend on either $\sigma^2_{\varepsilon}$ (the variance of the random error) or $\sigma^2_{\mu}$ (the variance of the random effect) or $\sigma^2_x$ (the variance of the time-varying covariate $x_{it}$).

\end{theorem}

\begin{theorem}
\label{thm: eveness_SEL}
Suppose that $N$, $T$, $\widetilde{\alpha}$, $1-\alpha$, $\psi$ and $\rho$ are fixed.
When $\sigma_{\varepsilon}$ and  $\sigma_{\mu}$ are replaced by any of the pairs of estimators listed in Appendix B,
the unconditional scaled expected length given in \eqref{SEL}
is an even function of $\tau \in (-1,1)$.
\end{theorem}


Suppose that $\widehat{\sigma}_{\varepsilon}$ and $\widehat{\sigma}_{\mu}$ are the usual unbiased estimators of $\sigma_{\varepsilon}$ and $\sigma_{\mu}$, respectively.  Consider the case that $N = 100$, $T=3$, $1 - \alpha = 0.95$ and $\widetilde{\alpha} \in \{0.05, 0.50\}$.  We find $c_{\text{min}}$ by minimizing the coverage probability over $\tau \in [0, 1)$, $\psi \in (0, \infty)$ and $\rho \in [0, 0.8]$, since it is reasonable to expect that $\rho \le 0.8$ in practice.
For given $\widetilde{\alpha}$ and $\rho$, we define min SEL and max SEL to be the scaled expected length minimized and maximized,
respectively, with respect to $\tau \in [0, 1)$ and $\psi \in (0, \infty)$.
For each $\widetilde{\alpha} \in \{0.05, 0.50\}$ and $\rho \in \{0, 0.2, 0.4, 0.6, 0.8\}$, we have computed min SEL and max SEL.
 This information is shown in Table 1.  From Table 1 it is clear that, for this example, $J_{\text{adj}}(\widehat{\sigma}_{\varepsilon})$ is the shorter interval on average.  The minimum scaled expected length is a decreasing function of $\rho$ because as $\rho$ increases, $\widehat{w}$ decreases.

\medskip

\begin{table}[h!]
\begin{center}
\begin{tabular}{ l c c c c c c c c c c c c }
\hline
$\widetilde{\alpha}$ & 0.05 & & &  &  &  & 0.50 & & & & & \\ \cline{2 - 6} \cline{8 - 12}
$\rho$ & 0 & 0.2 & 0.4 & 0.6 & 0.8 & & 0 & 0.2 & 0.4 & 0.6 & 0.8 & \\ \hline
min SEL & 4.06 & 3.67 & 3.23 & 2.70 & 1.99 & & 1.65 & 1.57 & 1.49 & 1.38 & 1.24 & \\
max SEL & 4.88 & 4.88 & 4.87 & 4.89 & 4.89 & & 1.81 & 1.81 & 1.81 & 1.81 & 1.81 &\\
\hline
\end{tabular}
\caption{Scaled expected lengths, both minimized (min SEL) and maximized (max SEL) over $\tau$ and $\psi$, for given values of $\widetilde{\alpha}$ and
$\rho$.}
\end{center}
\end{table}



\noindent {\sl Remark:} \
For the analysis of longitudinal data it is usually more appropriate to consider $G$ to be the matrix where the $(i,j)$'th element is $\rho^{|i-j|}$ (first order autoregression). Theorems \ref{thm: CovProb_DependOnPsi}, \ref{thm: eveness}, \ref{thm: SEL_DependOnPsi} and \ref{thm: eveness_SEL} are true for $G$ defined in this way.
Details on how to define $\tau$ in this case are provided in the supplementary material.

\section{The coverage probability when random error and random effect variances are assumed known}


\noindent 
When $\sigma_{\varepsilon}$ and $\sigma_{\mu}$ are known, the confidence interval resulting from the two-stage procedure is denoted by $K(\sigma_{\varepsilon}, \sigma_{\mu})$.
In this section we describe how the coverage probability of $K(\sigma_{\varepsilon}, \sigma_{\mu})$, conditional on $x$,  can be computed exactly using the bivariate normal distribution.
The results of this section are used in Section 4 to find control variates (used for variance reduction) for the estimation by simulation of $P(\beta \in K(\widehat{\sigma}_{\varepsilon}, \widehat{\sigma}_{\mu}))$ and the scaled expected length (given by \eqref{SEL}).



Let $P\big(\beta \in K(\sigma_{\varepsilon}, \sigma_{\mu}) \, \big| \, x \big)$ denote the coverage probability of
$K(\sigma_{\varepsilon}, \sigma_{\mu})$, conditional on $x$.
Observe that $P\big(\beta \in K(\sigma_{\varepsilon}, \sigma_{\mu}) \, \big| \, x \big)$ is equal to
\begin{align}
\label{CondCPKnown1}
\notag
&P \Big(\beta \in I(\sigma_{\varepsilon},\sigma_{\mu}), \, H(\sigma_{\varepsilon},\sigma_{\mu}) \le z_{1-\widetilde{\alpha}/2}^2 \, \Big| \, x \Big)
+ P \Big(\beta \in J(\sigma_{\varepsilon}), \, H(\sigma_{\varepsilon},\sigma_{\mu}) > z_{1-\widetilde{\alpha}/2}^2 \, \Big| \, x \Big) \\
&= P \big ( |g_I| \le z_{1-\alpha/2}, \, |h| \le z_{1-\widetilde{\alpha}/2} \, \big| \, x \big)
+ P \big ( |g_J| \le z_{1-\alpha/2}, \, |h| > z_{1-\widetilde{\alpha}/2} \, \big| \, x \big),
\end{align}
where
$g_I = \big(\widehat{\beta}(\psi)-\beta \big)/\big(\text{Var}_0(\widehat{\beta}(\psi)|x) \big)^{1/2}$,
$g_J = (\widetilde{\beta}_W-\beta)/\big(\text{Var}(\widetilde{\beta}_W|x)\big)^{1/2}$ and
$h = (\widetilde{\beta}_W - \widetilde{\beta}_B)/\big(\text{Var}(\widetilde{\beta}_W|x) + \text{Var}_0(\widetilde{\beta}_B|x)\big)^{1/2}$.
By the law of total probability, \eqref{CondCPKnown1}
is equal to the sum of $(1-\alpha)$ and
\begin{equation}
\label{CondCPKnown2}
P \big ( |g_I| \le z_{1-\alpha/2}, \, |h| \le z_{1-\widetilde{\alpha}/2} \, \big| \, x \big)
- P \big ( |g_J| \le z_{1-\alpha/2}, \, |h| \le z_{1-\widetilde{\alpha}/2} \, \big| \, x \big).
\end{equation}
The first and second terms in this expression are determined by the conditional
distributions of the random vectors $(g_I, h)$ and $(g_J, h)$,
respectively. Theorem \ref{known coverage} gives these distributions, which requires the introduction of
$p^2(x) = \SSB/\text{Var}(\overline{x}_i)$, where $\text{Var}(\overline{x}_i)$ is given in Appendix A.
The following theorem is stated by Kabaila et al 2015, together with an outline proof. A detailed proof
is proved in the supplementary material for the present paper.

\begin{theorem}
\label{known coverage}

Conditional on $x$, $(g_I, h)$ and $(g_J, h)$ have bivariate normal distributions, where $E(g_J \, | \, x) =  0$,
${\rm Var}(g_J \, | \, x) = 1$,
\begin{align*}
&E(g_I \, | \, x) =  \displaystyle{\frac{\tau \psi p(x)}{\big(q(\psi, T) + q^2(\psi, T)/r(x)\big)^{1/2}}}, \;
{\rm Var}(g_I \, | \, x) = 1 - \displaystyle{\frac{\tau^2 \psi^2}{q(\psi, T)  + q^2(\psi, T) / r(x)}}, \\
\\
&E(h \, | \, x) =  \displaystyle{\frac{-\tau \psi p(x)}{(r(x) + q(\psi,T))^{1/2}}}, \;
{\rm Var}(h \, | \, x) = 1 - \displaystyle{\frac{\tau^2 \psi^2 }{r(x) + q(\psi,T)}}, 
\end{align*}
\begin{align*}
&{\rm Cov}(g_I, h \, | \, x) = \displaystyle{\frac{\tau^2 \psi^2}{\big(q(\psi,T) r(x)+ q^2(\psi,T)\big)^{1/2} \big(1 + q(\psi,T)/r(x)\big)^{1/2}}} \\
\\
&\text{and} \ \ \ {\rm Cov}(g_J, h \, | \, x) = \displaystyle{\frac{1}{\big(1 + q(\psi,T)/r(x) \big)^{1/2}}}.
\end{align*}

\end{theorem}

Thus, when $\sigma_{\varepsilon}$ and $\sigma_{\mu}$ are known,
$P \big(\beta \in K(\sigma_{\varepsilon}, \sigma_{\mu})\, \big| \, x \big)$
can be found easily by evaluation of the bivariate normal cumulative distribution function
in the expression \eqref{CondCPKnown2}.
Similarly to Theorem \ref{thm: CovProb_DependOnPsi},
this probability
is determined by $N$, $T$, $x$, $\widetilde{\alpha}$, $1-\alpha$, $\psi$, $\rho$ and $\tau$.
Note that the dependence on $\rho$ is through $p(x)$.
Also, similarly to Theorem \ref{thm: eveness}, $P(\beta \in K(\sigma_{\varepsilon}, \sigma_{\mu})|x)$ is an even function of $\tau \in (-1,1)$.  These results may be proved using similar, but much simpler, arguments to those used in the proofs of Theorems \ref{thm: CovProb_DependOnPsi} and \ref{thm: eveness} which are given in the supplementary material.

\section{Simulation methods, including the use of variance reduction, when the random error and random \newline effect variances are unknown}

\noindent In Section 3 we described how to find the coverage probability of the confidence interval resulting from the two-stage procedure,
conditional on $x$ when $\sigma_{\varepsilon}$ and $\sigma_{\mu}$ are known, using the bivariate normal distribution.
In the practically important case that $\sigma_{\varepsilon}$ and $\sigma_{\mu}$ are replaced by estimators, the results of Section 3 allow us to find control variates (for variance reduction) for the estimation by simulation of both the coverage probability and the scaled expected length.
In Section 4.1 we describe the estimation by simulation of the coverage probability and in Section 4.2 we describe the estimation by simulation of the scaled expected length.
The simulation methods described in this section apply to any of the pairs of estimators $(\widehat{\sigma}_{\varepsilon}, \widehat{\sigma}_{\mu})$ listed in Appendix B.

We consider the model \eqref{model} and choose the intercept $a = 0$, the parameter of interest $\beta = 0$ and the
values for  $N$, $T$, $\widetilde{\alpha}$, $1 -\alpha$, $\sigma^2_{\varepsilon}$, $\sigma^2_{\mu}$ and $\sigma^2_{x}$.
Of course, by Theorem \ref{thm: CovProb_DependOnPsi} and Theorem \ref{thm: SEL_DependOnPsi}, the coverage probability and the scaled expected length
do not depend on either $a$, $\beta$ or $\sigma^2_{x}$ and depend on $\sigma^2_{\varepsilon}$ and $\sigma^2_{\mu}$ only through
$\psi=\sigma_{\mu}/\sigma_{\varepsilon}$.
Our simulation methods consist of $M$ independent simulation runs.
On the $k$'th simulation run ($k = 1, \dots, M$), we generate observations of the $\varepsilon_{it}$'s and $(\mu_i, x_{i1}, \dots, x_{iT})$'s using the assumptions made in Section 2, i.e. the $\varepsilon_{it}$'s are i.i.d. $N(0, \sigma^2_{\varepsilon})$ and the $(\mu_i, x_{i1}, \dots, x_{iT})$'s are i.i.d. with a multivariate normal distribution with mean $0$ and covariance matrix \eqref{CovMatrixMuiXit}.
As noted in Section 2, in simulations $x$ is determined by $u$, which we write as $x = x(u)$.
 Let $x^k$  and $u^k$ denote the observed values of $x$ and $u$, respectively, for the $k$'th simulation run.

\subsection{Estimating the coverage probability by simulation}

For the observed values in the $k$'th simulation run,
we compute the following three quantities.
The confidence interval resulting from the two-stage procedure, when $\sigma_{\varepsilon}$ and $\sigma_{\mu}$ are assumed known, denoted by $K_k(\sigma_{\varepsilon}, \sigma_{\mu})$.
The confidence interval resulting from the two-stage procedure,
when $\sigma_{\varepsilon}$ and $\sigma_{\mu}$ are estimated by $\widehat{\sigma}_{\varepsilon}$ and $\widehat{\sigma}_{\mu}$,
respectively, denoted by $K_k(\widehat{\sigma}_{\varepsilon}, \widehat{\sigma}_{\mu})$.
The coverage probability of $K(\sigma_{\varepsilon}, \sigma_{\mu})$, conditional on $x^k$, when $\sigma_{\varepsilon}$ and $\sigma_{\mu}$ are assumed known, is
$P \big(\beta \in K(\sigma_{\varepsilon}, \sigma_{\mu}) \, \big| \, x^k \big)$. Note that this conditional coverage probability is computed exactly using the bivariate normal distributions given in Theorem \ref{known coverage}.

Let $\CP = P(\beta \in K(\widehat{\sigma}_{\varepsilon}, \widehat{\sigma}_{\mu}))$, the coverage probability of $K(\widehat{\sigma}_{\varepsilon}, \widehat{\sigma}_{\mu})$.
Now define the unbiased estimator
\begin{equation*}
\widehat{\CP} = \frac{1}{M} \sum_{k=1}^M \mathcal{I} \big(\beta \in K_k(\widehat{\sigma}_{\varepsilon}, \widehat{\sigma}_{\mu}) \big)
\end{equation*}
of \CP. This is the usual ``brute-force'' simulation estimator of $\CP$.
We estimate the variance of this estimator by noting that it is a binomial proportion.
Let $\CPK = P \big(\beta \in K(\sigma_{\varepsilon}, \sigma_{\mu}) \big)$, the coverage probability of $K(\sigma_{\varepsilon}, \sigma_{\mu})$, when $\sigma_{\varepsilon}$ and $\sigma_{\mu}$ are assumed known. Now define the unbiased estimator
\begin{equation*}
\widehat{\CPK} = \frac{1}{M} \sum_{k=1}^M \mathcal{I} \big(\beta \in K_k(\sigma_{\varepsilon}, \sigma_{\mu}) \big)
\end{equation*}
of \CPK.
By the \textsl{double expectation theorem}, $\CPK = E_x \big(P \big(\beta \in K(\sigma_{\varepsilon}, \sigma_{\mu}) \, \big| \, x \big) \big)$.
Thus another unbiased estimator of $\CPK = P(\beta \in K(\sigma_{\varepsilon}, \sigma_{\mu}))$ is
\begin{equation*}
\widetilde{\CPK} = \frac{1}{M} \sum_{k=1}^M P \big(\beta \in K(\sigma_{\varepsilon}, \sigma_{\mu}) \, \big| \, x^k \big),
\end{equation*}
which is a much more accurate estimator of $\CPK$  than $\widehat{\CPK}$.

Define the \textsl{control variate} $\widehat{\CPK} - \widetilde{\CPK}$, which has expected value zero. The simulation-based unbiased estimator of
$\CP = P \big(\beta \in K(\widehat{\sigma}_{\varepsilon}, \widehat{\sigma}_{\mu}) \big)$
that employs variance reduction using this control variate,
is
\begin{equation*}
\widetilde{\CP} = \widehat{\CP} - \left (\widehat{\CPK} - \widetilde{\CPK} \right).
\end{equation*}
We expect that the correlation between $\widehat{\CP}$ and $\widehat{\CPK}$ will be close to 1.
Since $\widetilde{\CPK}$ is a much more accurate estimator of $\CPK$  than $\widehat{\CPK}$,
we expect that the correlation between $\widehat{\CP}$ and the control variate $\widehat{\CPK} - \widetilde{\CPK}$
will also be close to 1. Note that
\begin{equation*}
\widetilde{\CP}
= \frac{1}{M} \sum_{k=1}^M
\Big( \mathcal{I}\big(\beta \in K_k(\widehat{\sigma}_{\varepsilon}, \widehat{\sigma}_{\mu}) \big)
-  \mathcal{I}\big(\beta \in K_k(\sigma_{\varepsilon}, \sigma_{\mu}) \big)
+ P\big(\beta \in K(\sigma_{\varepsilon}, \sigma_{\mu}) \, \big| \, x^k \big) \Big).
\end{equation*}
We estimate the variance of this estimator by noting that it is an average of i.i.d. random variables.

We evaluate the efficiency gain of using $\widetilde{\CP}$ to estimate the coverage probability $\CP$ over $\widehat{\CP}$,
as follows.  Let $\widehat{\T}$ and $\widetilde{\T}$ denote the times taken to carry out $M$ simulation runs when we estimate $\CP$ by
$\widehat{\CP}$ and $\widetilde{\CP}$, respectively.
The efficiency
of the control variate estimator $\widetilde{\CP}$ relative to the ``brute-force'' estimator $\widehat{\CP}$ is
\begin{equation*}
\frac{\widehat{\T}}{\widetilde{\T}} \frac{\text{Var}(\widehat{\CP})}{\text{Var}(\widetilde{\CP})}.
\end{equation*}
The larger this relative efficiency, the greater the gain in using the control variate estimator $\widetilde{\CP}$, by comparison with using the ``brute-force'' estimator $\widehat{\CP}$.
To give an example of the efficiency gained by using $\widetilde{\CP}$ compared to $\widehat{\CP}$, when estimating $\CP$, we set $\rho = 0$, $N= 100$, $T=3$, $\tau = 0$, $\psi = 1/3$, $\alpha = \widetilde{\alpha} = 0.05$ and $M = 10,000$.  We obtain $\widehat{\T} = 179.37$ seconds, $\widetilde{T} = 211.51$ seconds, $\text{Var}(\widehat{\CP}) = 5.613591\times10^{-6}$ and $\text{Var}(\widetilde{\CP}) = 1.39591\times 10^{-6}$.  The time ratio is $\widehat{\T}/\widetilde{\T} = 0.848045$ and the variance ratio is $\text{Var}(\widehat{\CP})/\text{Var}(\widetilde{\CP}) = 4.92597$, so the efficiency of $\widetilde{\CP}$ relative to $\widehat{\CP}$ is approximately 4.17.  In other words, it would take approximately 4.17 times as long to compute the ``brute-force" estimator with the same accuracy as the control variate estimator.

We also use \textsl{common random numbers} to create smoother plots of the estimated coverage probability, as a function of $\lambda$.
The estimates of the coverage probability are computed for an equally-spaced grid of values of $\lambda$.
On the $k$'th simulation run we generate an observation of $(\mu_i, x_{i1}, \dots, x_{iT})$ 
from observations of the random numbers $u_{i1}, \dots, u_{iT}$ and $v_i$.
So, on the $k$'th simulation run, for each value of $\lambda$ in the grid, we use the same random numbers that are used to generate the observations of the $\varepsilon_{it}$'s and
the $(\mu_i, x_{i1}, \dots, x_{iT})$'s.  These observations are then used to construct our simulation-based estimate of $\CP$. Therefore on the $k$'th simulation run, for each value of $\lambda$, we have an estimate of the coverage probability using the same random numbers.

\subsection{Estimating the scaled expected length by simulation}

When estimating \eqref{SEL} by simulation we consider the expected values in the numerator and denominator seperately.
We begin with the expected value in the numerator of \eqref{SEL}.
Let $\textsc{LK}^{\dag} = (\widehat{\sigma}_{\varepsilon} / \sigma_{\varepsilon}) \left( \widehat{w}^{1/2} {\cal I}(\mathcal{B}) + {\cal I}({\cal B}^c) \right) \left( \sum_{i = 1}^N \sum_{t=1}^T (u_{it} - \overline{u}_i)^2 \right)^{-1/2}$ and let $\textsc{LK}^{\dag}_k$ be the value of $\textsc{LK}^{\dag}$ when $u = u^k$.  An unbiased simulation estimator of the expected value of $\textsc{LK}^{\dag}$ that does not use any variance reduction technique is
\begin{equation*}
\widehat{\textsc{LK}^{\dag}} = \frac{1}{M} \sum_{k=1}^M \textsc{LK}_k^{\dag}.
\end{equation*}
Now let
$\textsc{LKK}^{\dag} = \left( w^{1/2} {\cal I}({\cal C}) + {\cal I}({\cal C}^c) \right) \left( \sum_{i = 1}^N \sum_{t=1}^T (u_{it} - \overline{u}_i)^2 \right)^{-1/2}$,
where ${\cal C}$ is the statement $ -z_{1-\widetilde{\alpha}/2} \leq h \leq z_{1-\widetilde{\alpha}/2} $.
Also let $\textsc{LKK}^{\dag}_k$ be the value of $\textsc{LKK}^{\dag}$ when $u = u^k$.  Then $E(\textsc{LKK}^{\dag} \, | \, u)$ is equal to
\begin{align*}
&E\left(   w^{1/2} {\cal I}({\cal C}) + {\cal I}({\cal C}^c)  \, \big| \, u\right) \left( \sum_{i = 1}^N \sum_{t=1}^T (u_{it} - \overline{u}_i)^2 \right)^{-1/2}\\
& \indent = \left( w^{1/2} P({\cal C} \, | \, x) + P({\cal C}^c \, | \, x) \right) \left( \sum_{i=1}^N \sum_{t=1}^T (u_{it} - \overline{u}_i)^2 \right)^{-1/2},
\end{align*}
where $x = x(u)$.  
We compute $P({\cal C} \, | \, x)$ using Theorem \ref{known coverage}.
An unbiased simulation estimator of the expected value of $\textsc{LK}^{\dag}$ that makes use of a control variate for variance reduction is
\begin{equation*}
\widetilde{\textsc{LK}}^{\dag} = \frac{1}{M} \sum_{k=1}^M \left( \textsc{LK}_k^{\dag} - \left( \textsc{LKK}_k^{\dag} - E\left( \textsc{LKK}^{\dag} \, \big| \, u^k \right) \right) \right).
\end{equation*}

Next we will consider the expected value in the denominator of \eqref{SEL}.  Let $\textsc{LJ}^{\dag} = \left( \widehat{\sigma}_{\varepsilon} / \sigma_{\varepsilon} \right) \left( \sum_{i=1}^N \sum_{t=1}^T (u_{it} - \overline{u}_i)^2 \right)^{-1/2}$ and let $\textsc{LJ}^{\dag}_k$ be the value of $\textsc{LJ}^{\dag}$ when $u = u^k$.  An unbiased simulation estimator of the expected value of $\textsc{LJ}^{\dag}$ that does not use any variance reduction technique is
\begin{equation*}
\widehat{\textsc{LJ}^{\dag}} = \frac{1}{M} \sum_{k=1}^M \textsc{LJ}_k^{\dag}.
\end{equation*}
Now let $\textsc{LJK}^{\dag} = \left( \sum_{i=1}^N \sum_{t=1}^T (u_{it} - \overline{u}_i)^2 \right)^{-1/2}$ and let $\textsc{LJK}^{\dag}_k$ be the value of $\textsc{LJK}^{\dag}$ when $u = u^k$.  Note that $\sum_{i=1}^N \sum_{t=1}^T (u_{it} - \overline{u}_i)^2 \sim \chi^2_{N(T-1)}$.  It follows from this that $E\left(\textsc{LJK}^{\dag}\right) = 2^{-1/2} \left( \Gamma \left( \left( N(T-1) - 1\right)/2 \right) / \Gamma \left( \left( N(T-1) \right)/2 \right)\right)$.
An unbiased simulation estimator of the expected value of $\textsc{LJ}^{\dag}$ that makes use of a control variate for variance reduction is
\begin{equation*}
\widetilde{\textsc{LJ}}^{\dag} = \frac{1}{M} \sum_{k=1}^M \left( \textsc{LJ}_k^{\dag} - \left( \textsc{LJK}^{\dag}_k - 2^{-1/2} \frac{ \Gamma \left( (N(T-1)-1)/2 \right)}{\Gamma \left( (N(T-1))/2 \right)} \right) \right).
\end{equation*}

Therefore, a simulation estimator of the scaled expected length that makes use of control variates for variance reduction is
\begin{equation*}
\frac{z_{1-\alpha/2}}{\Phi^{-1}((c^* + 1)/2)} \frac{ \widetilde{\textsc{LK}}^{\dag}}{\widetilde{\textsc{LJ}}^{\dag}}.
\end{equation*}
We estimate the variance of $\widehat{\textsc{LK}}^{\dag}$, $\widetilde{\textsc{LK}}^{\dag}$, $\widehat{\textsc{LJ}}^{\dag}$ and $\widetilde{\textsc{LJ}}^{\dag}$ by noting that each estimator is an average of i.i.d. random variables.  Using R we can assess the relative efficiency of the estimators that use control variates for variance reduction with their ``brute-force'' counterparts using a similar method to that discussed in Section 4.1.
For example, for known quantities $N = 100$, $T = 3$, $\widetilde{\alpha} = 0.05$ and unknown quantities $\tau = 0$, $\rho = 0$, $\sigma_{\epsilon} = 1$, $\sigma_{\mu} = 1$ and $\sigma_x = 1$, and for $M = 1000$ independent simulation runs, we find that the efficiency of $\widetilde{\textsc{LK}}^{\dag}$ relative to $\widehat{\textsc{LK}}^{\dag}$ is approximately 1 and the efficiency of $\widetilde{\textsc{LJ}}^{\dag}$ relative to $\widehat{\textsc{LJ}}^{\dag}$ is approximately 2.
However, if we increase $\rho$ to 0.8 (leaving the other parameters unchanged), the efficiency of $\widetilde{\textsc{LK}}^{\dag}$ relative to $\widehat{\textsc{LK}}^{\dag}$ increases to approximately 3.
%
Hence, estimating the scaled expected length using control variates is well justified.

Common random numbers are also used for the estimation by simulation of the scaled expected length.  This is done in a similar way to the method described at the end of Section 4.1.

\noindent \textsl{Remark:} \ In this paper we consider the two-stage procedure when the inference of interest is a confidence interval for the slope parameter.  As one would expect from the duality between hypothesis tests and confidence intervals, our results have important implications when the subsequent inference is a hypothesis test for the slope parameter.  These implications are described in detail in the supplementary material.


\section{Conclusion}


\noindent The Hausman pretest is an example of preliminary statistical (i.e. data based) model selection. Other examples include model selection by minimizing a criterion such as the Akaike Information Criterion or the Bayesian Information Criterion. The effects of preliminary statistical model selection on confidence intervals can range from the benign to the very harmful, depending on the class of models under consideration, the known aspects of the model, the parameter of interest and the model selection procedure employed (Kabaila, 2009 and Kabaila and Leeb, 2006). In other words, each case needs to be considered individually on its merits.
Our results show that for the small levels of significance (such as 5\% or 1\%) of the Hausman pretest commonly used in applications,
the minimum coverage probability of the confidence interval for the slope parameter with nominal coverage probability $1-\alpha$ can
be far below nominal. The methodology that we have described makes it easy to assess, for a wide variety of circumstances, the effect of the
Hausman pretest on the minimum coverage probability of this confidence interval. An interesting finding is that if we increase the significance
level of the Hausman pretest to, say, 50\% then this minimum coverage probability is much closer to the nominal coverage $1-\alpha$
for a wide range of parameters.
However, regardless of the level of significance of the Hausman pretest, this interval is wider on average than the interval based on the fixed effects model which is adjusted to have the same minimum coverage.
Therefore, the Hausman prestest should not be used in practice to choose between the random effects model and fixed effects model for clustered or longitudinal data when the subsequent inference of interest is a confidence interval for the slope parameter.


\baselineskip=18pt

\section*{Appendix A. Definition of the non-exogeneity \newline parameter $\boldsymbol{\tau}$}

\noindent It may be shown that the distribution of $\mu_i$ conditional on $(x_{i1}, \dots, x_{iT})$
is normal with mean
\begin{equation*}
\frac{\sigma_{\mu} \, \widetilde{\tau} \, T}{\big(1 + (T-1) \rho \big) \, \sigma_x} \, \overline{x}_i,
\end{equation*}
where $\overline{x}_i = T^{-1} \sum_{t=1}^T x_{it}$, and variance
\begin{equation*}
\sigma_{\mu}^2 \left(1 -\frac{\widetilde{\tau}^2 \, T}{1 + (T-1) \rho} \right).
\end{equation*}
This suggests that $\tau = \text{Corr}(\mu_i, \overline{x}_i)$ is a reasonable measure of the dependence between
$\mu_i$ and $(x_{i1}, \dots, x_{iT})$ i.e. that it is reasonable to designate $\tau$ as the non-exogeneity parameter.
It may be shown that $\text{Var}(\overline{x}_i) = \sigma_x^2 (1 + (T-1) \rho)/T$ and
$\text{Cov}(\mu_i, \overline{x}_i) = \widetilde{\tau} \, \sigma_{\mu} \, \sigma_x$.
Thus
\begin{equation*}
\tau = \widetilde{\tau} \left( \frac{T}{1+(T-1)\rho} \right)^{1/2}.
\end{equation*}
%
%
%



\section*{Appendix B. Description of the estimators of the \newline random error and random effect variances}

\noindent It has been suggested in the literature (see e.g. Hsiao, 1986 and Baltagi, 2005) that if a negative estimate of variance is observed then
one should do as Maddala and Mount (1973) suggest and replace this negative estimate by 0. We use this kind of approach to ensure
that $\widehat{\sigma}^2_{\varepsilon}$  is always positive and $\widehat{\sigma}^2_{\mu}$ is always nonnegative. This ensures that
the proofs of Theorems \ref{thm: CovProb_DependOnPsi}, \ref{thm: eveness}, \ref{thm: SEL_DependOnPsi} and \ref{thm: eveness_SEL} carry through for each of the three pairs of estimators that we consider in this paper. We
consider the following pairs of estimators of $\sigma_{\varepsilon}^2$ and $\sigma_{\mu}^2$:

\medskip

\noindent (1) \ The usual unbiased estimators. Define
\begin{equation*}
\widehat{\sigma}_{\varepsilon}^2 = \frac{1}{N(T-1)-1} \sum_{i=1}^N \sum_{t=1}^T r_{it}^2
\end{equation*}
and
$
\widehat{\sigma}^2_{\mu} = \max(0, \, \widetilde{\sigma}^2_{\mu}),
$
where
\begin{equation*}
\widetilde{\sigma}^2_{\mu} = \frac{1}{N-2} \sum_{i=1}^N \overline{r}_i^2 - \frac{1}{NT(T-1)-T} \sum_{i=1}^N \sum_{t=1}^T r_{it}^2.
\end{equation*}
The $r_{it}$'s are the OLS residuals from model \eqref{consistent_est_beta_model} and the $\overline{r}_i$'s are the OLS residuals from model \eqref{average_over_t_model}.
Note that $\widetilde{\sigma}^2_{\mu}$ is an unbiased estimator of $\sigma^2_{\mu}$ only for $\tau=0$.

\medskip

\noindent (2) \ Hsiao's (1986) maximum likelihood estimators $\widehat{\sigma}^2_{\varepsilon}$ and
$\sigma^2_{\mu}$. We assume, of course, that the maximum likelihood estimator is obtained by maximizing the log-likelihood
function subject to the parameter constraints $\sigma_{\varepsilon}^2 \ge 0$ and $\sigma_{\mu}^2 \ge 0$.

 \medskip

\noindent (3) \   Wooldridge's (2002) estimators. Define
$
\widehat{\sigma}^2_{\varepsilon} = \max( -\epsilon \, \widetilde{\sigma}^2_{\varepsilon}, \, \widetilde{\sigma}^2_{\varepsilon})
$
where $\epsilon$ is a very small positive number and
\begin{equation*}
\widetilde{\sigma}_{\varepsilon}^{2} = \frac{1}{NT-K} \sum_{i=1}^N \sum_{t=1}^T \tilde{r}_{it}^2 - \frac{1}{NT(T-1)/2 - K} \sum_{i=1}^N \sum_{t=1}^{T-1} \sum_{s=t + 1}^T \tilde{r}_{it} \tilde{r}_{is} \, .
\end{equation*}
Also define
$
\widehat{\sigma}^2_{\mu} = \max(0, \, \widetilde{\sigma}^2_{\mu})
$
where
\begin{equation*}
\widetilde{\sigma}_{\mu}^2 = \frac{1}{NT(T-1)/2 -K} \sum_{i=1}^N \sum_{t=1}^{T-1} \sum_{s=t + 1}^T \tilde{r}_{it} \tilde{r}_{is} \, .
\end{equation*}
Here, the $\tilde{r}_{it}$'s are the residuals from pooled OLS estimation for the model \eqref{model}
and $K = 0$ (no d.o.f. correction) or $K = 2$ (d.o.f. correction).



\section*{Appendix C. Outline proofs of Theorems \ref{thm: SEL_DependOnPsi} and \ref{thm: eveness_SEL}}

\noindent Outline proofs of Theorems \ref{thm: CovProb_DependOnPsi}, \ref{thm: eveness} and \ref{known coverage} have been provided by Kabaila et al., 2015.  Here we provide outline proofs of Theorems \ref{thm: SEL_DependOnPsi} and \ref{thm: eveness_SEL}.  Detailed proofs of Theorems \ref{thm: CovProb_DependOnPsi} -- \ref{known coverage} can be found in the supplementary material.
Theorems \ref{thm: SEL_DependOnPsi} and \ref{thm: eveness_SEL} hold for the three  test statistics given by Hausman and Taylor (1981) and the three pairs of estimators described in Appendix B. For the sake of brevity, the outline proofs of these results are given only for the Hausman test statistic $H(\widehat{\sigma}_{\varepsilon},\widehat{\sigma}_{\mu})$ and the unbiased estimators of $\sigma_{\varepsilon}$ and $\sigma_{\mu}$ described in Appendix B.


\subsection*{Proof of Theorem \ref{thm: SEL_DependOnPsi}}
Let $\widehat{h}$ denote the statistic $h$, when $\sigma_{\varepsilon}$ and $\sigma_{\mu}$ are replaced by the unbiased estimators $\widehat{\sigma}_{\varepsilon}$ and $\widehat{\sigma}_{\mu}$, respectively.  Then ${\cal B}$ is equivalent to the statement that $-z_{1-\widetilde{\alpha}/2} \leq \widehat{h} \leq z_{1-\widetilde{\alpha}/2}$.
Let $x_{it}^{\dag} = x_{it}/ \sigma_x$, $\varepsilon_{it}^{\dag} = \varepsilon_{it} / \sigma_{\varepsilon}$ and $\mu_i^{\dag} = \mu_i / \sigma_{\mu}$.  The joint distribution of the $\varepsilon_{it}^{\dag}$'s and the $(\mu_i, x_{i1}, \dots, x_{iT})$'s is determined by $\rho$ and $\tau$.  Now express $\widehat{h}$ and $\widehat{\sigma}_{\varepsilon} / \sigma_{\varepsilon}$ in terms of the $x_{it}^{\dag}$'s, $\varepsilon_{it}^{\dag}$'s, $\mu_i^{\dag}$'s and $\psi$.  Since $c_{\text{min}}$ and the joint distribution of the 
$u_{it}$'s do not depend on any parameters, it follows that \eqref{SEL} is determined by the known quantities $N$, $T$, $1 - \alpha$ and $\widetilde{\alpha}$, and unknown parameters $\psi$, $\rho$ and $\tau$.

\subsection*{Proof of Theorem \ref{thm: eveness_SEL}}
Assume that $(\mu_i, x_{i1}, \dots, x_{iT})$ has a multivariate normal distribution with mean 0 and covariance matrix \eqref{CovMatrixMuiXit}.  Then $\widetilde{\tau} = \tau \left( (1 + (T-1) \rho) / T \right)^{1/2}$.
By Theorem \ref{thm: SEL_DependOnPsi}, the scaled expected length is a function of $\tau$.  Let $x_{it}^* = -x_{it}$ for $i=1, \dots, N$ and $t=1, \dots, T$.  For $\tau = d$, $(\mu_i, x_{i1}, \dots, x_{iT})$ has a multivariate normal distribution with mean 0 and covariance matrix \eqref{CovMatrixMuiXit}, where $\widetilde{\tau} = d \left( (1 + (T-1) \rho) / T \right)^{1/2}$.  For $\tau = -d$, $(\mu_i, x^*_{i1}, \dots, x^*_{iT})$ has the same distribution.  It can be shown that the scaled expected length is the same function of the $x_{it}^*$'s, $\varepsilon_{it}$'s and $\mu_i$'s as it is of the $x_{it}$'s, $\varepsilon_{it}$'s and $\mu_i$'s.  Hence the scaled expected length is an even function of $\tau$.

\newpage

\begin{center}
{\large\bf SUPPLEMENTARY MATERIAL}
\end{center}

\begin{description}

\item[Supplementary Appendix:] A remark on the two-stage procedure when the inference of interest is a hypothesis test and a remark on the choice of the scaling $\lambda = N^{1/2} \tau$.  This supplementary appendix also includes detailed proofs of Theorems 1 -- 7 and the definition of the non-exogeneity parameter $\tau$ in the case of first order autoregression. (PDF file)

%
%

\end{description}

%
\section*{References}

\rf Ajmani, V.B., 2009. Applied Econometrics Using the SAS system. John Wiley, Hoboken, N.J.

\smallskip

\rf Baltagi, B.H., 2005. Econometric Analysis of Panel Data, 3rd edition. John Wiley \& Sons, Ltd.

\smallskip

\rf Bedard, K., Deschenes, O., 2006.  The long-term impact of military service on health: Evidence from work war II and Korean war veterans.  American Economic Review 96, 176-194.

\smallskip

\rf Bloningen, B.A., 1997. Firm-specific assets and the link between exchange rates and foreign direct investment.
American Economic Review 87, 447--465.

\smallskip

\rf Croissant, Y., Millo, G., 2008. Panel data econometrics in R: The plm package. Journal of Statistical Software, 27(2), 1-43.

\smallskip

\rf Ebbes, P., Bockenholt, U., Wedel, M., 2004. Regressor and random-effects dependencies in multilevel models. Statistica Neerlandica 58, 161--178.

\smallskip

\rf Gardiner, J.C., Luo, Z., Roman, L.A., 2009.  Fixed effects, random intercepts and GEE: What are the differences? Statistics in Medicine 28, 221--239.

\smallskip

\rf Gaynor, M., Seider, H., Vogt, W.B., 2005.  The volume-outcome effect, scale economies, and learning-by-doing.  American Economic Review 95, 243-247.

\smallskip

\rf Griffiths, W. E., Hill, R. C., Lim, G., 2012. Using EViews for Principles of Econometrics. Danvers, MA: John Wiley \& Sons.

 \smallskip

\rf Guggenberger, P., 2010. The impact of a Hausman pretest on the size of a hypothesis test: The panel data case.
Journal of Econometrics 156, 337--343.

\smallskip

\rf Hastings, J.S., 2004. Vertical relationships and competition in retail gasoline markets: Empirical evidence from contract changes in Southern
California. American Economic Review 94, 317--328.

\smallskip

\rf Hsiao, C., 1986. Analysis of Panel Data. Cambridge University Press, Cambridge.

\smallskip

\rf Hausman, J.A., 1978. Specification tests in econometrics. Econometrica 46, 1251--1271.

\smallskip

\rf  Hausman, J.A., Taylor, W.E., 1981.  Panel data and unobservable individual effects.  Econometrica 49, 1377--1398.

\smallskip

\rf  Jackowicz, K., Kowalewski, O., Kozlowski, L., 2013.  The influence of political factors on commercial banks in Central European countries. Journal of Financial Stability 9, 759--777.



\smallskip

\rf Kabaila, P., 2009. The coverage properties of confidence regions after model
selection. International Statistical Review 77, 405--414.

\smallskip

\rf Kabaila, P., Leeb, H., 2006. On the large-sample minimal coverage probability of confidence intervals after model selection. Journal of the American Statistical Association 101, 619--629.

\smallskip

\rf Kabaila, P., Mainzer, R., Farchione, D., 2015.  The impact of a Hausman pretest, applied to panel data, on the coverage probability of confidence intervals.  Economics Letters 131, 12--15.

\smallskip

\rf Maddala, G.S., Mount, T.D., 1973.  A comparative study of alternative estimators for variance component models used in econometric applications.  Journal of the American Statistical Association 68, 324--328.

\smallskip

\rf Mann, V., De Stavola, B.L., Leon, D.A., 2004.  Separating within and between effects in family studies: an application to the study of blood pressure in children. Statistics in Medicine 23, 2745--2756.

\smallskip

\rf Rabe-Hesketh, S., Skrondal, A., 2012. Multilevel and Longitudinal Modeling Using Stata, 3rd edition. Stata Press, Texas.

\smallskip

\rf  Wooldridge, J. M., 2002.  Econometric Analysis of Cross Section and Panel Data.  MIT Press, Cambridge.

%
%
%
\end{document}